# Diffusion in translucent media

Zhou Shi[1,2] and Azriel Z. Genack[1*]

[1]*Department of Physics, Queens College and Graduate Center of the City University of New York, Flushing, New York 11367, USA*

[2]*Chiral Photonics Inc. 26 Chapin Road, Pine Brook, NJ 07058*

**Diffusion is the result of repeated random scattering. It governs a wide range of phenomena from Brownian motion, to heat flow through window panes, neutron flux in fuel rods, dispersion of light in human tissue, and electronic conduction. It is universally acknowledged that the diffusion approach to describing wave transport fails in translucent samples thinner than the distance between scattering events such as are encountered in meteorology, astronomy, biomedicine and communications. Here we show in optical measurements and numerical simulations that the scaling of transmission and the intensity profiles of transmission eigenchannels have the same form in translucent as in opaque media. Paradoxically, the similarities in transport across translucent and opaque samples explain the puzzling observations of suppressed optical and ultrasonic delay times relative to predictions of diffusion theory well into the diffusive regime.**

## Introduction

Einstein showed that microscopically-visible particles buffeted by stochastic molecular forces perform a random walk that can be described by the diffusion equation once the initial motion of particles is randomized[1]. The diffusion approach also describes the transport of classical and quantum waves in multiply scattering media[2-20]. Waves entering a static disordered sample interfere to produce a wavelength-scale speckled pattern of energy or particle density that is a unique fingerprint of the wave interaction with the disordered sample. When such patterns are averaged over a large ensemble of statistically equivalent samples, a smoothed profile of energy density results that is a solution of the diffusion equation[6]. The diffusion approach is assumed to fail, on time scales shorter than the scattering time[9] and on length scales smaller than the transport mean free path, $\ell^1$, in which the particle direction is randomized. On these scales, it is assumed that transport can only be described by a detailed accounting of radiative transfer within the sample[2,20].

The transmission of waves through a disordered material is fully characterized by the transmission matrix, $t$, whose elements $t_{ba}$ are the field transmission coefficients between complete sets of $N$ orthogonal propagating channels on each side of the sample[21-32]. For an incident field in channel $a$, $E_a$, the transmitted field in channel $b$, $E_b$, can be expressed as the sum of the coherent field, with the same intensity pattern as the incident field, and a random field, which is uncorrelated with $E_a$, $E_b = E_{coherent} + E_{random} = \langle t_{ba} \rangle E_a \delta_{ab} + \delta E_b$. Here $\langle \cdots \rangle$ represents the average over random sample configurations and $\delta_{ab} = 1$ for a=b, and 0 otherwise.

In this article, we explore the relationship between wave propagation in translucent and diffusive samples. A widely-held view is that transport in these regimes is totally dissimilar. True, diffusion is built from a series of random ballistic steps. However, the wave retains a degree of spatially coherence during each step, whereas multiply scattered waves are randomized with vanishing correlation across the sample. As a result, many characteristics of transport are totally different in these two regimes, as is illustrated in the next section, and propagation is described using different formalisms.

Here we show that, notwithstanding the stark differences between transport in translucent and opaque samples, the underlying structure of transport is strikingly similar. The scaling of transmission and the energy density inside a random medium illuminated by random waveforms have identical forms. The energy density inside the sample falls linearly and extrapolates to zero at the same distance beyond the sample in both regimes. At the same time, the average energy density profiles in the interior of specific transmission eigenchannels have nearly identical forms. We show that the source of these similarities is the correlation within the transmission matrix, which leads to characteristic repulsion between transmission eigenvalues on all length scales. The surprisingly short dwell time observed in the crossover from ballistic to diffusive propagation is shown to be a consequence of the diffusive form of the energy density profile for the perfectly transmitting eigenchannel.

**Results**

**Coherent versus randomized waves in translucent and opaque samples.**

The dominance of coherent or ballistic light in optically thin samples and of incoherent multiply-scattered light in opaque samples is illustrated in the recursive Green's function simulations[33] shown in Fig. 1. Simulations are carried out for a scalar wave of wavelength $\lambda_0 = 650$ nm propagating through a two-dimensional strip with reflecting sides along its length. A random segment of length $L$ is sandwiched between regions of dielectric constant unity. The disordered region is divided into square elements with sides of length $\lambda_0/2\pi = 103.5$ nm and dielectric function $\varepsilon(x, y) = 1+\delta\varepsilon(x, y)$ with $\delta\varepsilon(x, y)$ selected randomly from a uniform distribution in the range -0.2 and 0.2. The strip of width $W = 5.2$ μm supports $N = 16$ propagating waveguide modes. The $n = 1\ldots16$ waveguide modes have transverse profiles $\phi_n(y) \sim \sin(k^n_y y)$ with transverse components of the *k*-vectors $k^n_y = n\pi/W$ and longitudinal speed $v_n = ck^n_x/k$, where *c* is the speed of light (Details of the simulations are given in the Methods section).

In translucent samples, the transmission coefficient of coherent flux is of order unity, $|\langle t_{nn} \rangle|^2 \sim 1$, as seen in Fig. 1a. In contrast, the coherent flux in diffusive media is exponentially small, as seen in Fig. 1b. The coherent flux, $|\langle t_{nn}(L) \rangle|^2$, falls exponentially with sample length *L* at different rates for each of the *N* waveguide modes (Fig. 1c). However, the coherent flux falls at a single rate in the time domain, $1/\tau_s$, as seen in the inset of Fig. 1c. This yields the scattering mean free time and so the scattering mean free path, $\ell_s = c\tau_s = 27.2$ µm. Since the scale of the scattering element is much smaller than the wavelength, and fluctuations in $\varepsilon$ are small, $\ell_s$ is expected to be nearly equal to $\ell$ [5].

The average delay time in transmission, $t_D$, which equals the average of the delay relative to the incident wave of the transmission channels weighted by the corresponding transmission eigenvalue, is shown in Fig. 1d (Supplementary Note 4). $t_D$ scales linearly for translucent samples and, in the thinnest samples, is equal to the average delay over all waveguide modes for a sample without disorder, $t_B = \langle L/v_n \rangle \equiv L/v_+$. Thus $v_+$ represents the average longitudinal component of velocity of a random incident wave. For the samples studied in simulations, $v_+ = 0.70c$. $t_D$ approach quadratic scaling only for $L \gg \ell$.

**Scaling of optical transmission**

Since waves are largely coherent in translucent samples and randomized in diffusive media, one might expect the total transmission to scale differently in these regimes. Surprisingly, however, measurements of total optical transmission, which includes both the scattered and unscattered waves, were found to be in accord with diffusion theory down to sample lengths of $L \sim 2\ell$[8,13,15].

We explore wave propagation on still shorter length scales with $L \ll \ell$ to discover whether there is a lower limit in thickness below which the diffusion model fails. We note that computer simulations of the scaling of transmission of the portion of light that has been scattered at least once can be described by diffusion theory, even for $L \ll \ell$[14]. Here, however, we consider the full transmitted flux including light that has not been scattered, as is generally the case in measurements of transmission.

For $L \gg \ell$, the scaling of average transmission of an incident beam is found by solving the diffusion equation with the impact of the boundary incorporated phenomenologically[13]. For a single incident channel $a$, the ensemble average of total transmission is, $\langle T_a \rangle = (z_{p,a}+z_b)/(L+2z_b)$[13] (Supplementary Equation 7), where $z_{p,a}$ is the effective penetration depth of radiation in channel $a$ at which radiation is randomized and $z_b$ is the distance beyond the sample boundary in which the intensity within the sample extrapolates to zero. The model is solved for a randomized source at a depth $z_{p,a}$ with strength equal to the intensity that enters the sample. Surprisingly, the above expression is in excellent agreement with measurements down to $L = 2\ell$ [13]. But one might not expect this model to apply to samples thinner than the penetration depth, since the effective source would then fall beyond the output boundary of the sample.

To explore transport in the crossover from ballistic to diffusive propagation, we measure the scaling of optical transmission through a dilute latex colloid contained in two wedge-shaped sample holders with different wedge angles. A normally incident laser beam is softly focused on the front of the sample while the transmitted light is collected in an integrating sphere (Details of the optical measurements are given in Methods section.) The thickness of the sample through which light passes is varied by translating the sample vertically perpendicular to the vertex of the wedge. The inverse of total transmission for the channel $a$ corresponding to the normally incident beam, $1/\langle T_a \rangle$, is seen in Fig. 2a to increase linearly with $L$ over the combined range of thicknesses in the two wedged samples of from $L = 20$ μm to 2.5 mm. From the distance beyond the sample of $2z_b$ at which $1/\langle T_a \rangle$ extrapolates to zero and the value of $2z_b/(z_{p,a}+z_b)$ to which $1/\langle T_a(L) \rangle$ extrapolates at $L=0$, we obtain $z_b = 0.93$ mm and $z_{p,a} = 0.76$ mm. This gives $\ell \sim 0.94$ mm[13]. The linearity of measurements of $1/\langle T_a(L) \rangle$ from $0.05\ell$ to $2.7\ell$ shows that transmission follows the diffusion model even for $L \ll \ell$. Agreement of the scaling of transmission in the translucent regime with diffusion theory is also found in simulations in random 2D waveguides of the inverse of the total transmission averaged over all incident channels, $1/\langle T_a \rangle_a$, shown in Fig. 2b. Thus, despite the differences in propagation between translucent and opaque samples shown in Fig. 1, the expressions for the scaling of total transmission for a single incident channel (Fig. 2a) and for the average over all incident channels (Fig. 2b) apply equally in translucent and opaque media.

**Energy density distribution inside opaque and translucent media**

For diffusive waves, the flux though the sample is proportional to the spatial derivative of the energy density within the sample. It is of interest therefore to compare energy density profiles in samples thinner and thicker than $\ell$. Diffusion theory predicts a linear falloff of the average energy density with depth into a sample illuminated with a mixture of all incident waveguide modes. This is precisely what is found in the simulations shown in Fig. 2c for translucent as well as diffusive samples. Moreover, we find that the energy density extrapolates to zero at the same distance, $z_b$ = 19.2 ± 0.2 µm from the output surface for both opaque and translucent samples. This value of $z_b$ is in accord with the value found in simulations of the scaling of transmission shown in Fig. 2b of $z_b$ = 19.1 ± 0.1 µm.

In Fig. 2c, we plot $W(x)$, the energy density integrated over the transverse direction at a depth $x$ averaged over random configurations and incident waveguide modes. $W(x)$ is normalized so that at it is equal to the average transmission coefficient through the sample at $x = L$, $W(L) = \langle T / N \rangle = u(L)v_+$. The transmittance $T$ is the sum over all channel-to-channel flux transmission coefficients, $T = \sum_{a,b=1}^{N} |t_{ba}|^2$, while $u(x)$ is the average energy density of a wave for unit incident flux.

The flux through a sample is given by Fick's first law of diffusion, $\langle T_a \rangle_a = -D \dfrac{du(x)}{dx}$, where $D$ is the diffusion coefficient. In two dimensional samples, $D$ is given by $v\ell/2$, where $v$ is the speed of the wave. Since $W(x)$ extrapolates to zero at a distance $z_b$ beyond the output surface of the sample, we can show that $\ell = 2z_b v_+/v$ (Supplementary Equation 4). This relation gives $\ell$ = 26.9 ± 0.3 µm which is close to the value of $\ell_s$ = 27.2 ± 0.2 found from Fig. 1c. Thus both transmission and the energy density within the sample are well described by diffusion theory even in translucent samples.

**Transmission eigenvalues**

The scaling of conductance and transmission in multiply scattering media can be expressed in terms of the transmission eigenvalues, $\tau_n$. These are the ensemble averages of the $N$ eigenvalues of the $N \times N$ Hermitian matrix product $tt^\dagger$, where $t^\dagger$ is the Hermitian conjugate of the transmission matrix $t$. The $\tau_n$ are indexed in order of decreasing transmission from $n = 1$ to $N$ and are proportional to the energy on the output surface of the sample; their sum gives the average transmittance, $\langle T \rangle = \sum_1^N \tau_n$. The scaling of transmission eigenvalues, and, hence of the transmittance or conductance, was described by Dorokhov[21] in terms of a set of auxiliary localization lengths, $\xi_n$, where, $\tau_n = 1/\cosh^2 x_n$ with $x_n = L/\xi_n$. For $L \gg \ell$. The $x_n$ scale linearly for $n < N/2$ with spacing, $x_{n+1}-x_n \equiv \Delta x = L/\xi$, where $\xi = N\ell$ is the localization length. For $n > N/2$, the $x_n$ increase somewhat more rapidly[24,26].

Though waves in translucent samples are not randomized, the transmission matrix can still be defined and the scaling of the $x_n$ can be computed in simulations in the translucent as well as the diffusive regime. We find a common structure for the $x_n$ with the $x_n$ remaining equally spaced for $n < N/2$, as shown in Fig. 3a. The structure persists even in the thinnest samples for which the spacing is no longer proportional to $L/N\ell$ (Supplementary Figure 3).

Another striking manifestation of universality is seen in the probability distributions of the normalized spacing between adjacent $x_n$, $s = (x_{n+1}-x_n)/\Delta x$ for $n < N/2$. The distributions shown in Fig. 3b collapse to a single curve corresponding to Wigner's surmise for the Gaussian orthogonal ensemble for eigenvalues of large random matrices[26]. This distribution, predicted for diffusive samples, is found to hold even for translucent samples. This reflects the universal repulsion between the $x_n$ seen in Fig. 3a and produces the same scaling law for transmission in translucent and diffusive samples.

**Transmission eigenchannels**

Since the similarity in the scaling of transmission in translucent and diffusive samples is related to the similarity in the $x_n$, and so the $\tau_n$, it is useful to explore whether there is a similarity in form between energy density of the transmission eigenchannels in translucent and diffusive media. This will determine the energy density inside the sample, and ultimately the delay time in transmission[34-39] (Supplementary Equation 10).

The transmission eigenchannels at the incident and output boundaries of the sample and the transmission eigenvalues are obtained from the singular value decomposition of the transmission matrix, $t$ [26]. The field within the sample for the $n^{th}$ transmission eigenchannel cannot be obtained from $t$, but is just the field generated in the interior of the sample by the incident waveform for the transmission eigenchannels. We will consider $W_n(x)$ or $W_\tau(x)$, the contribution to $W(x)$ of the $n^{th}$ transmission eigenchannel or the eigenchannel with transmission $\tau$, which are normalized so that on the output surface, $W_n(L) = \tau_n$ or $W_\tau(L) = \tau$. The average profile of energy density throughout the sample excited by a mix of all incident channels is, $W(x) = \sum_1^N W_n(x)/N,$ or equivalently an integral over the product of $W_\tau(x)$ and the probability density of $\tau$. To arrive at an expression for the functional form of the energy density profiles, it is useful to consider the scaling of the transmission eigenchannel profiles and to consider the profiles as functions of $x/L$, $W_\tau(x/L)$.

In diffusive samples, $W_\tau(x/L)$ can be written as the product of the profile of the completely transmitting eigenchannel with $\tau = 1$, $W_1(x/L)$, and a function $S_\tau(x/L)$, which is independent of $L/\ell$ and depends only on $\tau$, $W_\tau(x/L) = W_1(x/L)S_\tau(x/L)$[39]. $W_1(x/L)$ can be expressed as $1+F_1(x/L)$, where $F_1(x/L)= A(L/\ell)[4(x/L)(1-x/L)]$ is a solution of the diffusion equation with boundary conditions appropriate for perfect transmission[39]. $A(L/\ell)$ is the peak value of $F_1(x/L)$ at $x/L=1/2$. We show in Figs. 4a,b that when $F_1(x/L)$ is normalized by its peak value, the curves for translucent and diffusive media collapse to the function $4(x/L)(1-x/L)$. Thus the spatial structure of the perfectly transmitting eigenchannel is the same in translucent and diffusive media.

We present results for $S_\tau(x/L)$ for $L/\ell = 0.18$ for various values of $\tau$ in Fig. 4c. It has not been possible to derive the expression for $S_\tau(x/L)$ for diffusing waves from first principles. However, the expression for transmission eigenvalues $\tau_n$ in terms of $x_n = L/\zeta_n$ suggests a possible analytical expression for $S_\tau(x/L)$, which is in good agreement with the simulations in Fig. 4c. For a given value of $\tau$, the expression for $S_\tau(x/L)$ is an extension of Dorokhov's expression for $\tau_n$ on the surfaces of the sample into the interior[21]. The values of $S_\tau=W_\tau$ at $x=L$ and 0 of $\tau$ and $(2-\tau)$, respectively, are consistent with the expression, $S_\tau(x/L) = 2\tau\cosh^2((1-x/L)L/\xi')-\tau$, where $\tau$ is given by $1/\cosh^2(L/\xi')$. This expression matches the results of simulations in translucent samples for various values of $\tau$ shown in Fig. 4c. In diffusive samples, however, the expression above for $S_\tau(x)$ shows a systematic departure from simulations (Supplementary Figure 5). Agreement with simulations in diffusive samples is only obtained once an empirical function is added in the argument of the hyperbolic cosine in the expression above for $S_\tau(x)$ [39] (Supplementary Figure 6).

A complete description of propagation in random media requires the scaling of the energy density profiles of transmission eigenchannels and so the scaling of $W_1(x/L)$. The form of the energy density for the completely transmitting eigenchannel, $W_1(x/L) = 1+A(L/\ell)[4(x/L)(1-$ $x/L$)] does not change throughout the translucent and diffusive regimes as seen in Fig. 4a,b. To find the scaling of $W_1(x/L)$, it remains to find the scaling of $A(L/\ell)$. The variation of the peak value of $W_1(x/L)$ with $L/\ell$ is plotted in Fig. 5a and fit to the sum of a constant of unity and a linear term and a leading quadratic correction in $L/\ell$. The coefficient of the linear term is found to be 0.355.

Solving a generalized diffusion equation with flux at the output equal to the incident flux yields the peak value of $A(L/\ell) = v_+L/2v\ell$ (Supplementary Note 4). We have shown above that for our sample, the ratio of $v_+$ and $v$ is 0.7. This gives a linear contribution to $A(L/\ell)$ of 0.35, in agreement with the coefficient found in simulations. When $L$ approaches $\xi$, $A(L/\ell)$ is expected to increase more rapidly because coherent backscattering enhances the return of the wave to points in the medium[40]. Thus $W_1(x)$ is seen to be the sum of a constant "ballistic" term, a linear "diffusive" term, and "localization" correction that becomes important as $L$ approaches the localization length $N\ell$.

**Dwell times**

Measurements of optical[11,15,18,19,41] and ultrasound[16] pulsed transmission through random slabs show that on average photons arrive earlier than predicted by diffusion theory even in samples with $L > 5\ell$. The average delay time $t_D$ can also be determined from the transmission eigenvalues and energy density profiles of the transmission eigenchannels[36] (Supplementary Note 4). It can be expressed as the average delay time of the transmission eigenchannels $t_n$ weighted by the corresponding transmission eigenvalues, $\tau_n$, $t_D = \sum_1^N \tau_n t_n / \sum_1^N \tau_n$. The eigenchannel delay time is proportional to the energy stored within the sample so that $t_n \sim \int_0^L W_n(x)dx$ [36] (Supplementary Note 4).

In Fig. 5b, we plot $t_D$ and the delay time of the fully transmitting eigenchannel, $t_1$. Since the form of $S_\tau(x)$ is independent of $L/\ell$ for diffusive waves, the scaling of $t_D$ for $N > L/\ell > 1$ largely depends upon the scaling of $t_1$, which is given by the integral of $W_1(x)$ over the sample length. Only for $L/\ell = 2.65$ is the amplitude of the "diffusive" component of $W_1(x/L)$, equal to the value of the "ballistic" component, while the value of the integral of the diffusive term over the sample length only reaches that for the ballistic term for $L/\ell = 3.82$. In addition to the small slope of $A(L/\ell)$ vs. $L/\ell$ in thin samples, the dwell time increases slowly in thin samples because the superlinear increases of the integral of $W_1(x)$ (Supplementary Equation 10) is offset by the sublinear increases of the $t_n$ (Supplementary Figure 7). In contrast, for thicker samples, $\tau_n$ is typically small for channels $n>g$ so that low transmission eigenchannels do not contribute appreciably to $t_D$ (Supplementary Figure 7). For these reasons, the onset of diffusive scaling of the dwell time only begins when $L/\ell$ is substantially larger then unity. Thus, it is precisely the similarities in the functional form of characteristics of static transport between translucent and opaque samples which lead to reduced delay times relative to predictions of the diffusion model.

The shorter delay time in transmission relative to diffusion theory, which is found from the energy density inside the sample as a sum of eigenfunctions of the diffusion equation[11], limits the time in which the wave can spread in the transverse direction and so results in a reduced width of the transverse profile of intensity on the output surface in thin samples[13] and early times[18] relative to diffusion theory. In thicker strong scattering samples, observations of a halt in the transverse spread of the intensity profile on the output surface indicate that the wave is localized[42]. Though the present study has focused on longitudinal propagation in translucent and diffusive quasi-one dimensional samples, the evolution of the transverse intensity distribution with sample thickness in samples of any scattering strength can be studied in the slab geometry within the framework of transmission eigenchannels by decomposing a narrow incident beam into a sum of transmission eigenchannels

**Discussion**

A consistent picture of propagation in the crossover from ballistic to multiple scattering has long remained elusive. On the one hand, the scaling of transmission in samples hardly thicker than a mean free path still obeys diffusion theory, while on the other, the dwell time in samples up to several times the mean free path scale only slightly faster than linearly, as would be expected for waves following nearly ballistic trajectories. This work shows that the questions raised are even more perplexing since measurements of optical transmission are found to scale diffusively down to one fiftieth of the mean free path.

We show here that a description of the energy density and flow within random translucent and opaque systems emerges from the common statistics of the ratios of the sample length and eigenchannel localization lengths, $x_n = L/\xi_n$, and the intensity profiles of the associated transmission eigenchannels. Transmission is determined by the sum over transmission eigenvalues, which reflects the mutual repulsion of $x_n$, while the deviation of dwell time from diffusion theory is a consequence of the diffusive form of the energy density profiles of transmission eigenchannels even in translucent samples. The delay time for diffusive samples is largely determined by the profile of the fully transmitting transmission eigenchannel $W_1(x/L)$, which includes a factor which is the sum of a constant ballistic term, a diffusive term linear in $L/\ell$, and a leading-order localization correction which is quadratic in $L/\ell$. It is the small coefficient of the linear term relative to unity which is largely responsible for the slow approach to the quadratic scaling of $t_D$ associated with diffusion.

The delay time in reflection, which is of importance in optical or ultrasound diffuse tomography, can also be given in terms of the properties of transmission eigenchannels. Since the delay time of transmission eigenchannels is the same in reflection as in transmission[36] and the reflection coefficient in the $n^{th}$ transmission eigenchannel is $(1-\tau_n)$, the average delay time in reflection is $t_D^{\text{reflection}} = \sum_1^N (1-\tau_n) t_n / \sum_1^N \tau_n$ [36].

The work in this paper opens the door for study of many open issues. Among these are a fuller expression for the localization contribution to $W_1(x/L)$, not only the coefficient of the normalized function $F_1(x/L)/F_1(1/2)$, but also the deviation of this function from the diffusive form for localized waves. If propagation is primarily through single peaked localized states, one would expect that $F_1(x/L)/F_1(1/2)$ would narrow significantly since the intensity should be peaked within a localization length of the center of the sample for high maximal transmission[43]. But if the width of this function does not change appreciably, transport would then largely be through coupled localization centers, known as necklace states, in which the incident is coupled strongly through the sample[44]. Thus, the narrowing of the width of $F_1(x/L)/F_1(1/2)$ would indicate the dominance of the transport through either isolated states or necklace states for localized waves. The existence of both single peaked localized states and multiply peaked necklace states has been observed in layered media[45], single mode waveguides[46], natural materials[47], and can be created in multimode optical fiber with mode coupling[48]. It is also of great interest to explore the disposition of energy within thin anisotropic scattering media, of importance in biomedical research[49].

Obtaining the mean free path over the full range of opacity is also of importance in monitoring colloidal, micellar, or metallic nanoparticle concentrations, sedimentation, atmospheric conditions, and medical diagnostics. Since the scaling of transmission and time delay depend on $\ell$ and $z_b$ in different ways, the results presented here suggest that it should be possible to determine the mean free path in samples over a broad range of $L/\ell$. In future work the relationship between $\ell$ and $z_b$ in the presence of internal reflection will be determined in the regime of the crossover from translucent to multiply scattering samples. These results would, for example, provide a path towards quantitative monitoring of particulate concentrations in liquids or gases in sample with thickness of the order of the mean free path. The transport mean free path can also be obtained from the spacing of the $x_n$ in translucent samples, in which the measurements of the TM can be more complete since the number of coherence areas is relatively small in translucent media[30].

Recent developments of techniques for measuring the transmission matrix for imaging applications are relevant to both thin and thick scattering samples. A clearer picture of the connection between energy density and time delay in scattering are of importance in many approaches to imaging. For example, in medical imaging, different regions of a sample are probed in diffusing temporal field correlation spectroscopy[50] as the distance between the probe and source are changed, while different dwell times within the medium may be probed even for fixed spacing by utilizing correlation spectroscopy in the time domain[51]. These techniques are important in non-invasively monitoring blood flow and managing the delivery of oxygen to the brain.

**Methods**

**Numerical simulations of a scalar wave propagating**

The Green's function $G(\mathbf{r},\mathbf{r}')$ between arrays of points on the input surface $\mathbf{r} = (0, y)$ and at a depth $x$, $\mathbf{r}' = (x, y)$ can be obtained by solving the wave equation $\nabla^2 E(x,y) + k_0^2 \varepsilon(x,y) E(x,y) = 0$ on a square grid via the recursive Green's function method. To calculate the transmitted flux for various incident and output waveguide modes, the Green's function is expressed in terms of the basis of the waveguide modes, $t_{ba}(x) = \sqrt{v_b v_a} \int_0^W dy' \int_0^W dy\, \phi_b(y) \phi_a^*(y') G(\mathbf{r},\mathbf{r}')$, in which $v_a$ is the group velocity of the waveguide mode $a$, and $W$ is the width of the waveguide.

The incident wavefront $\mathbf{v_n}$ and outgoing filed $\mathbf{u_n}$ associated with the n$^{th}$ eigenchannel can be found using the singular value decomposition of the transmission matrix, $t = U\Lambda V^+$, where $\mathbf{u_n}$ and $\mathbf{v_n}$ are columns of the unitary matrix $U$ and $V$, respectively. $\Lambda$ is a diagonal matrix with elements $\sqrt{\tau_n}$. The field at a depth $x$ for an incoming eigenchannel in momentum space is found by multiplying the transmission matrix $t_{ba}(x)$ by $\mathbf{v_n}$. Summing the square of the coefficients over the $N$ waveguide modes yields the density of the flux at $x$. At the output surface, $x = L$, this gives $\tau_n$. The energy density $W_n(x)$ can then be obtained by dividing the density of the flux by the average speed $v_+$ of the wave propagating through the waveguide. The scaling of the total transmission shown in Fig. 2b was obtained by averaging over 5,000 sample configurations. $W_n(x)$ for $L = 5.2$ μm and 124.2 μm was averaged over 200,000 and 10,000 samples, respectively and the energy distributions for eigenchannels with a specific value of transmission $\tau$ are found by averaging the eigenchannel with transmission between $0.98\tau$ and $1.02\tau$. To find the scaling of the peak value of the $F_1(x)$, 500 sample realizations were averaged for each of the lengths of samples to ranging from 5.2 μm to 154.5 μm to yield the $\langle W_1(x) \rangle$. The profile of the fully transmitting eigenchannels for $\tau > 0.98$ was subsequently fitted with a parabolic function to give the peak value.

**Optical measurements of light propagation through a wedged random medium**

The scaling of total transmission is measured for a colloid of 0.17-μm-diameter polystyrene spheres in water at a volume fraction of ~ 0.003. An anionic surfactant was added to the colloidal suspension to prevent particle aggregation. The latex spheres and surfactant were obtained from Polysciences. The colloid is placed in two wedged sample holders made from microscope slides meeting at vertex angles of $\theta_{\text{wedge}} = 0.86°$ and $5.88°$. Polished glass and aluminum wedges were used as spacers between the slides. The sides of the assembly were sealed with wax. The normally incident beam of light at 532 nm is weakly focused on the incident face of the sample. The sample is translated perpendicular to the vertex line in steps of 1 mm after each measurement of transmission. The light spreads to a spot on the output plane with diameter of order of $L$. Because the wedge angles are small, the variation in thickness $L$ of the colloid across the illuminated region of the sample is much smaller than the sample thickness $L$. The transmitted light is collected in a Labsphere integrating sphere.

**Data availability**

The authors declare that all data that support the findings of this study are available from Zhou Shi at zhoushi.qc@gmail.com upon reasonable request.

**References**

1. Einstein, A. *Investigations on the Theory of the Brownian Movement.* Courier Corporation, (1956).

2. Milne, E. A. Radiative equilibrium in the outer layers of a star. *Monthly Notices of the Royal Astronomical Society*, **81**, 361-375 (1921).

3. Morse, P. M. & Feshbach, H. *Methods of Theoretical Physics.* (McGraw-Hill, New York, 1953).

4. Cercignani, C. *The Boltzmann Equation and its Applications* (Springer, New York, 1988).

5. Ishimaru, A. *Wave Propagation and Scattering in Random Media* (Wiley-IEEE Press, 1999).

6. Van Rossum, M. C. W. & Nieuwenhuizen, Th. M. Multiple scattering of classical waves: microscopy, mesoscopy, and diffusion, *Rev. Mod. Phys.* **71,** 313-371 (1999).

7. Shapiro, B. Large intensity fluctuations for wave propagation in random media. *Phys. Rev. Lett.* **57**, 2168-2171 (1986).

8. Genack, A. Z. Optical Transmission in Disordered Media. *Phys. Rev. Lett.* **58**, 2043-2046 (1987).

9. Li, T., Kheifets, S., Medellin, D. & Raizen, M. G. Measurement of the instantaneous velocity of a Brownian particle. *Science* **328**, 1673-1675 (2010).

10. Lagendijk, A., Vreeker, R. & De Vries, P. Influence of internal reflection on diffusive transport in strongly scattering media, *Phys. Lett. A* **136**, 81-88 (1989).

11. Yoo, K. M., Liu, F. & Alfano, R. R. When does the diffusion approximation fail to describe photon transport in random media? *Phys. Rev. Lett.* **64**, 2647-2650 (1990).

12. Zhu, J. Pine, D. J. & Weitz, D. A. Internal reflection of diffusive light in random media. *Phys. Rev. A* **44**, 3948-3959 (1991).

13. Li, J. H., Lisyansky, A. A., Cheung, T. D., Livdan, D. & Genack, A. Z. Transmission and Surface Intensity Profiles in Random Media. *Europhys. Lett.* **22**, 675-680 (1993).

14. Durian, D. J. Influence of boundary reflection and refraction on diffusive photon transport, *Phys. Rev. E* **50**, 857-866 (1994).

15. Kop, R. H., de Vries, P., Sprik, R. & Lagendijk, A. Observation of anomalous transport of strongly multiple scattered light in thin disordered slabs. *Phys. Rev. Lett.* **79**, 4369-4372. (1997).

16. Zhang, Z. Q., Jones, I. P., Schriemer, H. P., Page, J. H., Weitz, D. A., and Sheng, P. Wave transport in random media: The ballistic to diffusive transition, *Phys. Rev. E* **60**, 4843-4850 (1999).

17. Zhao, L., Tian, C., Bliokh, Y. P. & Freilikher, V. Controlling transmission eigenchannels in random media by edge reflection. *Phys. Rev. B* **92**, 094203 (2015).

18. Pattelli, L., Mazzamuto, G., Wiersma, D. S., & Toninelli, C. Diffusive light transport in semitransparent media, *Phys. Rev. A* **94**, 043846 (2016).

19. Pattelli, L., Savo, R., Burresi, M., & Wiersma, D. S. Spatio-temporal visualization of light transport in complex photonic structures, *Light: Science & Applications* **5**, e16090 (2016).

20. Chandrasekhar, S. *Radiative Transfer* (Dover Publications, Inc. 1960, Cambridge, 2007).

21. Dorokhov, O. N. On the coexistence of localized and extended electronic states in the metallic phase. *Solid State Commun.* **51**, 381–384 (1984).

22. Imry, Y. Active transmission channels and universal conductance fluctuations. *Euro. Phys. Lett.* **1**, 249–256 (1986).

23. Mello, P. A., Pereyra, P. & Kumar, N. Macroscopic approach to multichannel disordered conductors. *Ann. Phys.* **181**, 290–317 (1988).

24. Pichard, J. L., Zanon, N., Imry, Y. & Stone, A. D. Theory of random multiplicative transfer matrices and its implications for quantum transport. *J. Phys. France* **51**, 587-609 (1990).

25. Pendry, J. B., MacKinnon, A. & Pretre, A. B. Maximal fluctuations - a new phenomenon in disordered systems. *Physica A: Statistical Mechanics and its Applications* **168**, 400-407 (1990).

26. Beenakker, C. W. J. Random-matrix theory of quantum transport, *Rev. Mod. Phys*. **69**, 731-808 (1997).

27. Popoff, S. M., Lerosey, G., Carminati, R., Fink, M., Boccara, A. C. & Gigan, S. Measuring the transmission matrix in optics: an approach to the study and control of light propagation in disordered media. *Phys. Rev. Lett*. **104**, 100601 (2010).

28. Shi, Z. & Genack, A. Z. Transmission eigenvalues and the bare conductance in the crossover to Anderson localization. *Phys. Rev. Lett.* **108**, 043901 (2012).

29. Kim, M. *et al*. Maximal energy transport through disordered media with the implementation of transmission eigenchannels. *Nat. Photon.* **6**, 583–587 (2012).

30. Goetschy, A. & Stone, A. D., Filtering Random Matrices: The Effect of Incomplete Channel Control in Multiple Scattering, *Phys. Rev. Lett.* **111**, 063901 (2013).

31. Gèrardin, B., Laurent, J., Derode, A., Prada, C. & Aubry, A. Full transmission and reflection of waves propagating through a maze of disorder, *Phys. Rev. Lett.* **113**, 173901 (2014).

32. Rotter, S. & Gigan, S, Light fields in complex media: Mesoscopic scattering meets wave control. *Rev. Mod. Phys.* **89,** 015005 (2017).

33. Baranger, H. U., DiVincenzo, D. P., Jalabert, R. A. & Stone, A. D. Classical and quantum ballistic-transport anomalies in microjunctions. *Phys. Rev. B* **44**, 10637-10675 (1991).

34. Avishai, Y. & Band, Y. One-dimensional density of states and the phase of the transmission amplitude. *Phys. Rev. B* **32**, 2674-2676 (1985).

35. Iannacone, G. General relation between density of states and dwell times in mesoscopic systems. *Phys. Rev. B* **51**, 4727–4729 (1995).

36. Davy, M, Shi, Z., Wang, J., Cheng, X. & Genack, A. Z. Transmission eigenchannels and the densities of states of random media. *Phys. Rev. Lett.* **114**, 033901 (2015).

37. Choi, W., Mosk, A. P., Park, Q. & Choi, W. Transmission eigenchannels in a disordered medium. *Phys. Rev. B* **83**, 134207 (2011).

38. Sarma, R., Yamilov, A. G., Petrenko, S. Bromberg, Y. & H. Cao H. Control of Energy Density inside a Disordered Medium by Coupling to Open or Closed Channels. *Phys. Rev. Lett.* **117**, 86803 (2016).

39. Davy, M., Shi, Z., Park, J., Tian, C. & Genack, A. Z. Universal structure of transmission eigenchannels inside opaque media. *Nat. Commun.* **6**, 6893 (2015).

40. Tian, C., Hydrodynamic and field-theoretic approaches to light localization in open media, *Physica E*, **49**, 124 (2013).

41. Badon, A., Li, D., Lerosey, G., Bocara, A., Fink, M., and Aubry, A., Spatio-temporal imaging of light transport in highly scattering media under white light illumination, *Optica*, **3**, 1160 (2016)

42. Sperling, T., Bührer, W., Aegerter, C., and Maret, G., Direct determination of the transition to localization of light in three dimensions, *Nat. Photonics*, **7**, 48–52 (2012)

43. Azbel, M. Y. Eigenstates and properties of random systems in one dimension at zero temperature. *Phys. Rev.* **28**, 4106-4125 (1983)

44. Pendry, J. B. Symmetry and transport of waves in 1D disordered systems. *Adv. Phys.* **43**, 61–542 (1994).

45. Bertolotti, J., Gottardo, S., Wiersma, D. S., Ghulinyan, M., and Pavesi, L., Optical necklace states in Anderson localized 1D systems. Phys. Rev. Lett., **94**, 113903 (2005).

46. Sebbah, P., Hu, B., Klosner, J. M., and Genack, A. Z., Extended quasimodes within nominally localized random waveguides, Phys. Rev. Lett., **96**, 183906 (2006)

47. Seung Ho Choi, S. H., Byun, K. M., & Kim, Y. L. Lasing interactions disclose hidden modes of necklaces states in the Anderson localized regime. *ACS Photonics*, DOI: 10.1021/acsphotonics.7b01110 (2017).

48. Xiong, W., Ambichl, P., Bromberg, Y., Redding, B., Rotter, S., and Cao, H., Spatiotemporal Control of Light Transmission through a Multimode Fiber with Strong Mode Coupling, Phys. Rev. Lett. , **117**, 053901 (2016).

49. Judkewitz, B., Horstmeyer, R., Vellekoop, I. M., Papadopoulos, I. N., and Yang, C., Translation correlations in anisotropically scattering media, Nat. Phys., **11**, 684 (2015).

50. Boas, D. A., Campbell, L. E. & Yodh, A. G. Scattering and imaging with diffusing temporal field correlations. *Phys. Rev. Lett.* **75**, 1855-1858 (1995).

51. Sutin, J., Zimmerman, B., Tyulmankov, D., Tamborini, D., Wu, K. C., Selb, J., Gulinatti, A., Rech, I., Tosi, A., Boas, D. A. & Franceschini, M. A. Time-domain diffuse correlation spectroscopy, Optica, **3**, 1006-1013 (2016).

**Acknowledgements**

We thank X. Cheng for help with the optical measurements and acknowledge useful discussions with V. A. Gopar, X. Cheng, M. Davy, and C. Tian. This research was supported by the National Science Foundation (DMR-BSF: 1609218).

**Author contributions**

Z.S. carried out the numerical simulations. A.Z.G. performed the optical measurements and the calculations. A. Z. G. and Z. S. wrote the manuscript.

**Competing interests:** Authors have no competing financial interest.

**Correspondence:**

Correspondence and requests for materials should be addressed to genack@qc.edu.

**Figures**

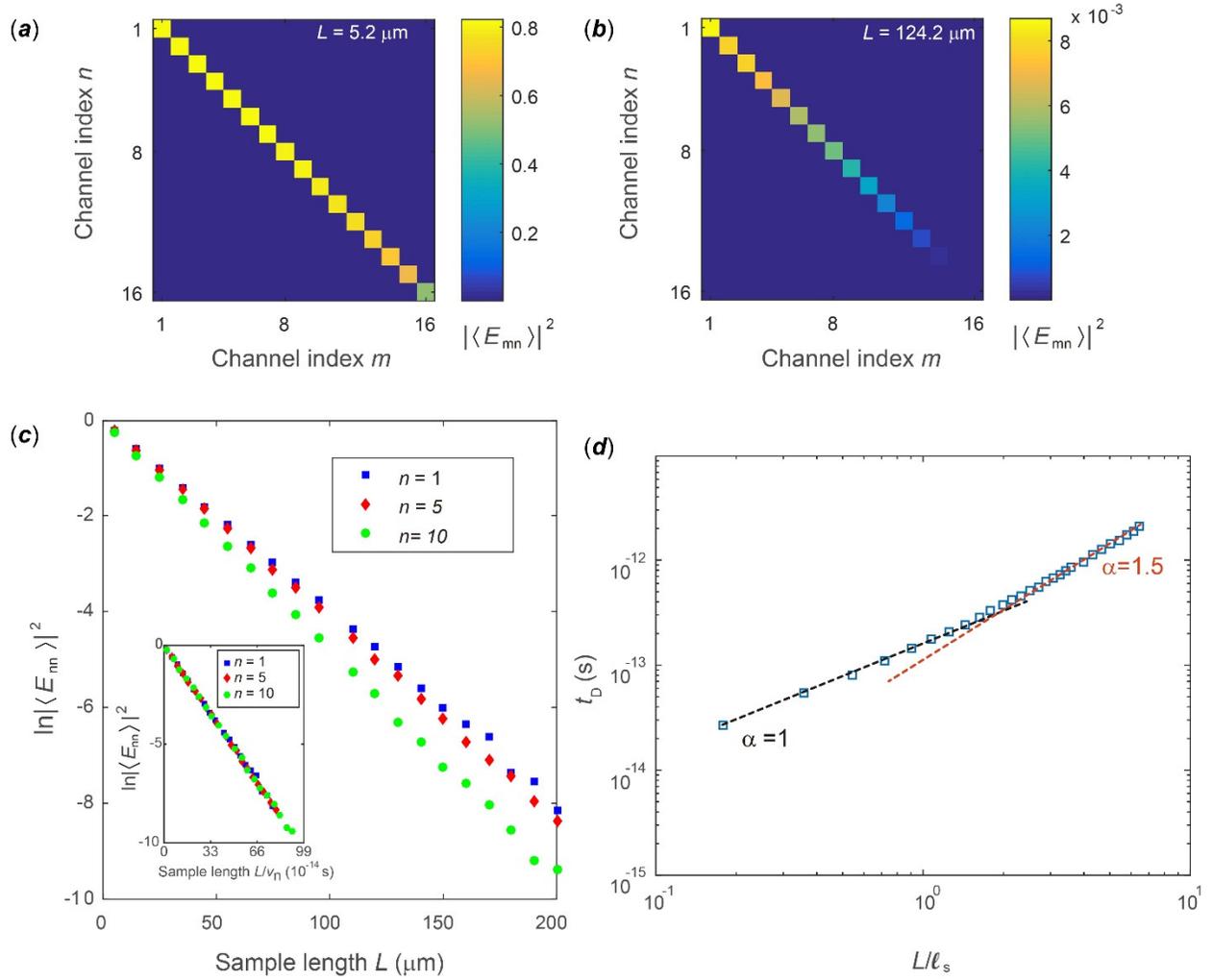

**Figure 1| Simulations of wave transmission in opaque and translucent samples. (a,b)** Average transmitted intensity $|\langle E_{mn}\rangle|^2$ for different incident and output waveguide modes. The coherent intensity, for m = n, is substantial for $L = 5.2$ µm and negligible for $L = 124.2$ µm. **(c)** Scaling of the coherent intensity. The inset shows the scaling of three incident waveguide modes with $n$ = 1,5,10. The variation of $|\langle E_{nn}\rangle|^2$ with time delay $L/v_n$ for these modes with longitudinal velocities $v_n$ collapses to a single curve and falls exponentially to give a scattering length of $\ell_s$ = 27.2±0.2 µm. The scattering length is given by, $\ell_s = c\tau_s$, where $\tau_s$ is the mean free time obtained

from the decay rate in the insert. **(d)** Log-log plot of the transmission delay time $t_D$ with sample length $L$. The dashed lines indicate different exponent $\alpha$ of the power law scaling. The transition from linearly scaling occurs for $L \sim \ell$. For $L \gg \ell$, the value of $\alpha$ is close to 2.

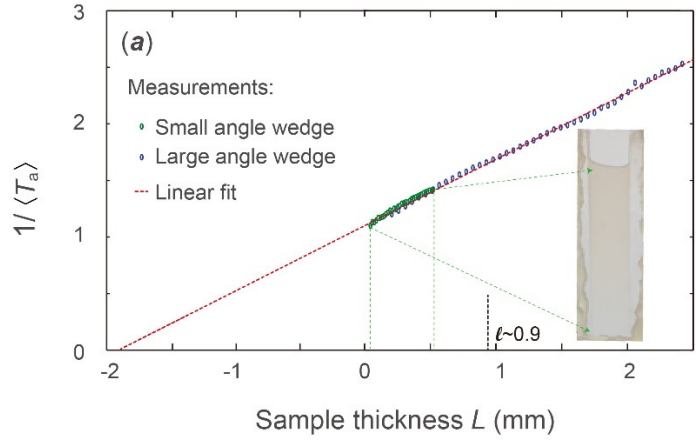

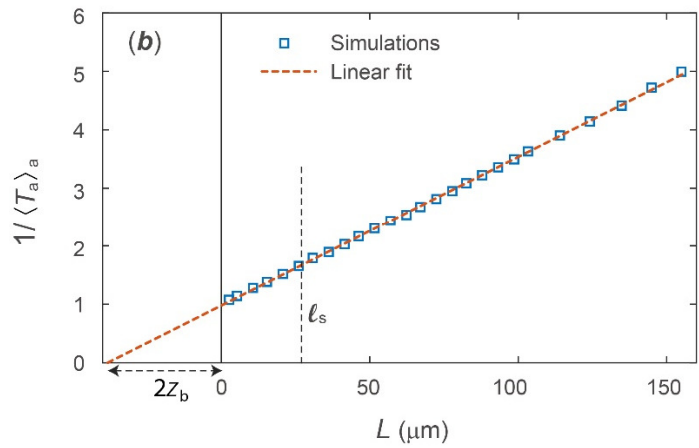

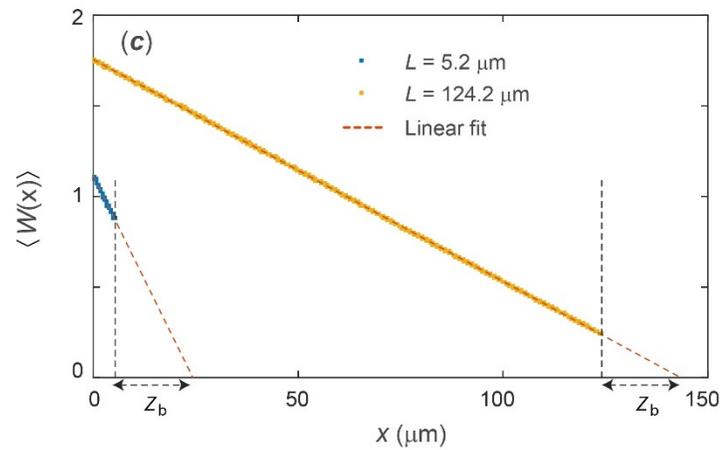

**Figure 2| Scaling of total transmission. (a)** Measurements of the scaling of the inverse of the optical transmission through a dilute suspension of 0.17-μm-diameter latex spheres in water. A photograph of the face of the translucent sample with a wedge angle of 0.86° is shown. The thicknesses at the beginning and end of the scan are indicated by the dashed green lines. The sides of the wedge are not shown because the microscope slides forming the faces of the sample are attached at their sides to a glass wedge with wax. The determination of $\ell \sim 0.9$ mm is discussed in the text. The value of $z_b$ is increased due to surface reflection at the air-glass interfaces. **(b)** Simulations of the scaling of the inverse of the total transmission averaged over all incident channels extrapolates to zero at $2z_b$, giving $z_b = 19.1 \pm 0.1$ μm. The vertical solid line indicates $L = 0$ μm and the vertical dashed line gives the value of $\ell_s$. **(c)** Results of simulations show a linear falloff of average energy density inside both translucent and multiple-scattering samples. The energy density extrapolates to zero beyond the sample boundary at the same distance $z_b = 19.2 \pm 0.2$ μm in both samples. The boundaries of the two samples are indicated by the dashed vertical lines.

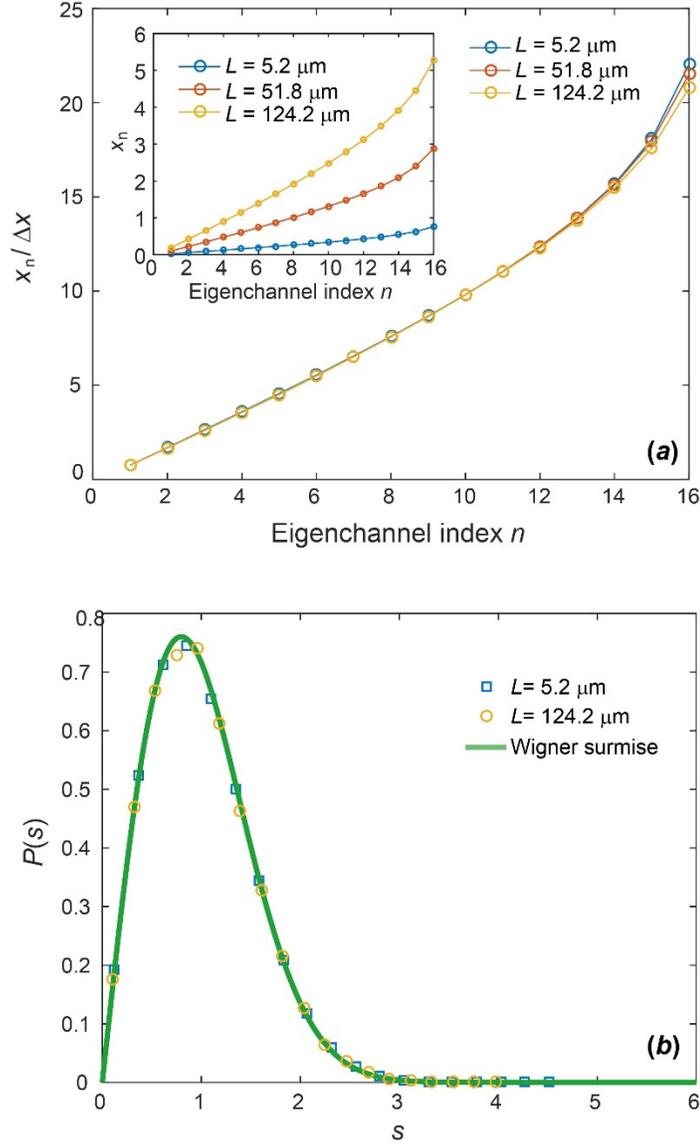

**Figure 3| Structure of the transmission eigenvalues.** The $x_n$ are determined by the transmission eigenvalues via the expression, $\tau_n = 1/\cosh^2 x_n$ with $x_n = L/\xi_n$. (**a**) The variation of the $x_n$ relative to the average spacing between them, $\Delta x$, are similar for different sample lengths, $L$. (**b**) The probability distribution of the spacing between the $x_n$ for $n < N/2$ in individual configurations is in accord with the Wigner surmise[26] for the eigenvalues of large random matrices for the Gaussian orthogonal ensemble, $P(s) = \frac{\pi}{2} e^{-\pi s^2/4}$, for both translucent and diffusive samples.

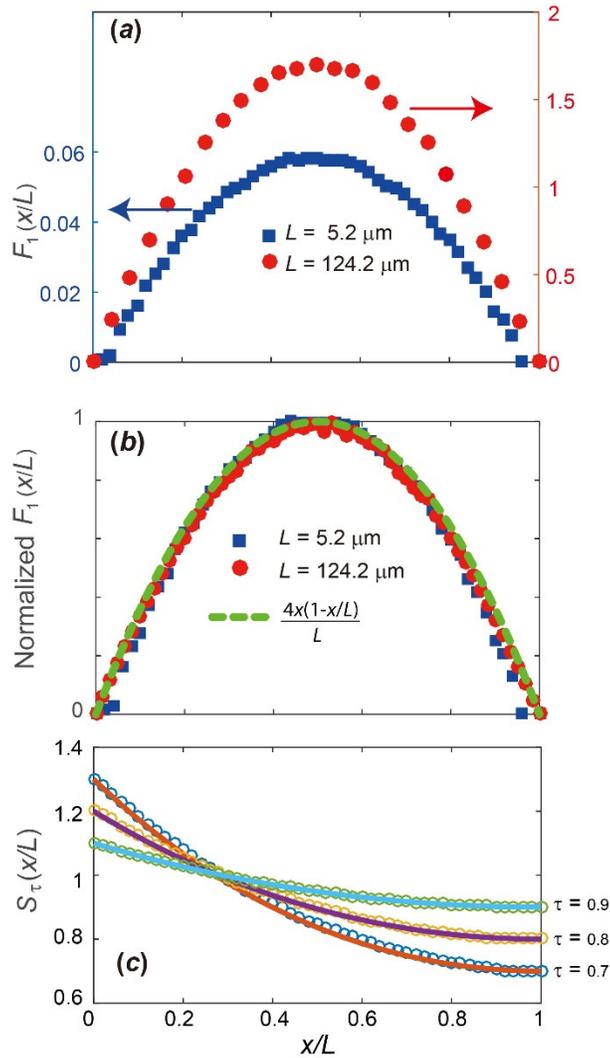

**Figure 4| Profiles of energy density in transmission eigenchannels.** (**a**) Profiles of completely transmitting eigenchannel in the translucent and diffusive regimes. (**b**) Profiles of $F_1(x/L) = W_1(x/L)-1$ normalized by its peak value in the center of the sample for translucent and diffusive samples collapse to $4x(L-x)/L^2$. (**c**) Comparison of $S_\tau(x/L)$ found from simulations compared with the expression, $S_\tau(x/L) = 2\tau\cosh^2((1-x/L)L/\xi')-\tau$, for values of $\tau = 0.7$, 0.8, and 0.9 in a sample with $L/\ell = 0.18$. $S_\tau(x/L)$ for small values of $\tau$ are not shown because small values occur so infrequently.

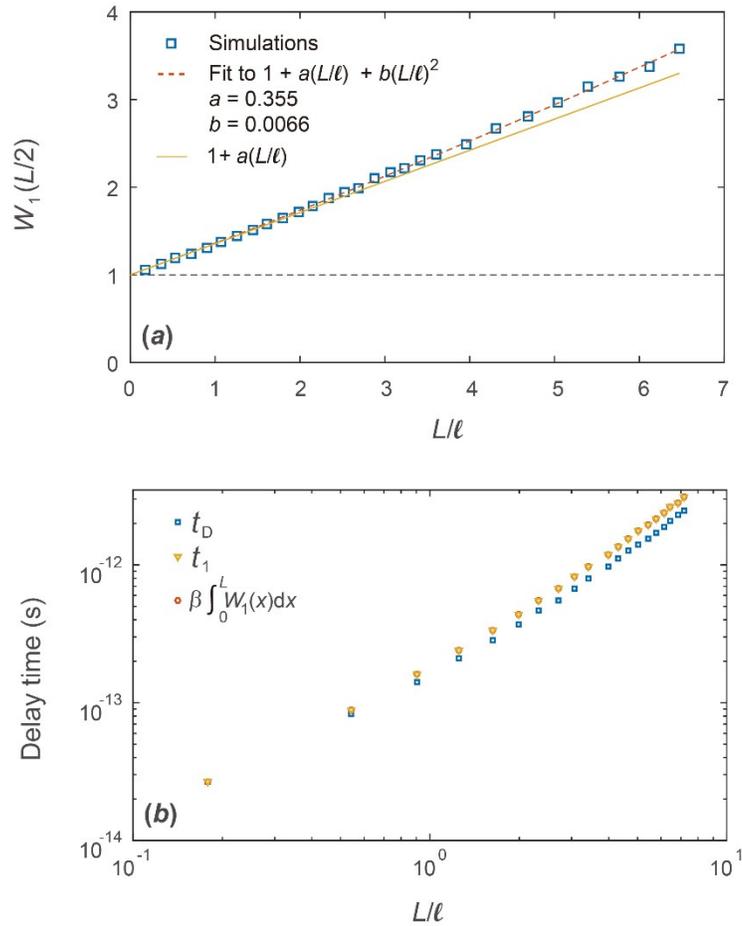

**Figure 5| Scaling of $W_1(L/2)$ and the delay time of the fully transmitting eigenchannel $t_1(L)$ and the average delay time $t_D(L)$** **(a)** The scaling of $W_1(L/2)$ (blue squares) is fit by a parabolic function $1 + a(L/\ell) + b(L/\ell)^2$. The fit gives $a = 0.355$ and $b = 0.0066$ (red dashed curve). The coefficient $a$ can be calculated using diffusion theory, while the quadratic term reflects enhanced delay due to incipient localization. The sum of the constant term of unity (black dashed line) and the linear term of $a(L/\ell)$ is shown as the yellow solid line. **(b)** The integral of $W_1(x)$, which is proportional to the delay time of the fully transmitting eigenchannel, is shown as the red circles

in the log-log plot of Fig. 5b. The delay time of the fully transmitting eigenchannel obtained from the composite phase derivative of the eigenchannel with respect to the frequency shift[37] is shown as the triangles in Fig. 5b (Supplementary Note 4). The scaling of $t_D$, shown as the blue squares, is similar to the scaling of $t_1$ for diffusive waves.

**Supplementary Information for "***Diffusion in translucent media***"**
**Shi et al.**

# Supplementary Information for "Diffusion in translucent media"
## Zhou Shi[1,2] and Azriel Z. Genack[1]

[1]Department of Physics, Queens College and Graduate Center of the City University of New York, Flushing, New York 11367, USA

[2]Chiral Photonics Inc. 26 Chapin Road, Pine Brook, NJ 07058

**Supplementary Note 1 - Spatial parameters in the diffusion model**

The flow of incoherent wave energy in an unbounded non-dissipative medium is described by the diffusion equation (*1-4*),

$$\frac{\partial u(\mathbf{r},t)}{\partial t} = -D\nabla^2 u(\mathbf{r},t) = Q(\mathbf{r},t) \tag{1}$$

Here $u(\mathbf{r}, t)$ is the energy density, $D = v\ell/d$ is the diffusion coefficient, $v$ is the transport velocity, $\ell$ is the transport mean free path, $d$ is the dimensionality, and $Q(\mathbf{r}, t)$ is a source function of incoherent waves. The diffusion equation only describes the evolution of energy density of fully randomized waves, however, the diffusion model can provide the evolution of energy density created by a coherent incident beam incident within a bounded sample by phenomenologically incorporating interactions at the interface (*1-4*). A coherent incident beam is replaced with a delta function source of incoherent radiation at a depth $z_p$ into the sample. For an incident beam with angle of refraction within the medium $\theta$, the penetration depth is $z_p = z_{p0}\cos\theta$, where $z_{p0}$ is the penetration depth for a normally incident beam (*4*). The boundary may be eliminated by solving the diffusion equation in an unbounded medium in which the linearly decaying intensity near the boundaries on either side of the source extrapolates to zero at a distance $z_b$ beyond the sample. This is illustrated in Supplementary Figure 1.

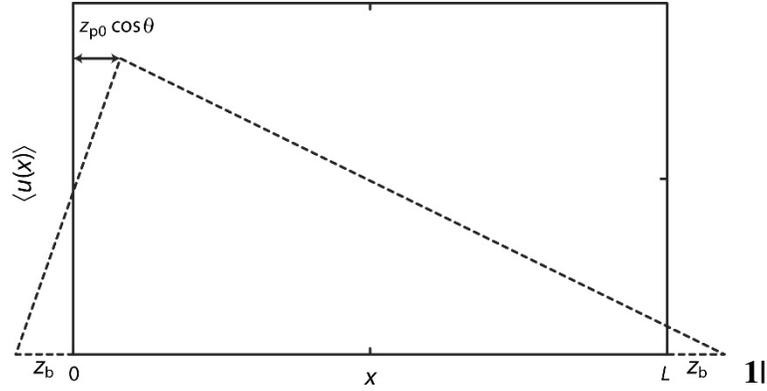

**Supplementary Figure 1 | Illustration of the diffusion model within a bounded random medium.** Interactions at the boundaries are incoporated using phenomenological lengths. An incoherent beam with refracted angle $\theta$ is replaced by an isotropic source at a depth $z_{p0}\cos\theta$ into the sample from which the energy density diffuses freely. The energy density near the boundaries extrapolates to zero at a distance $z_b$ beyond the sample's open surfaces.

In this model, the reflected flux, for unit incident flux in channel $a$, $\langle R_a \rangle$ is the flow to the left from the source at $z_{p0}\cos\theta$, while the transmitted flux, $\langle T_a \rangle$, is the flow to the right from the internal source. These fluxes are given by $-D\dfrac{du(x)}{dx}$, according to Fick' first law. The gradients of the energy density are proportional to the inverse of the distance from $z_p$ to the points beyond the sample at which the energy density extrapolates to zero, so that $\dfrac{\langle T_a \rangle}{\langle R_a \rangle} = \dfrac{z_{p0}\cos\theta + z_b}{L + z_b - z_p}$.

Together with the condition for conservation of energy, $\langle R_a \rangle + \langle T_a \rangle = 1$, this gives (4),

$$\langle T_a \rangle = \frac{z_{p0}\cos\theta + z_b}{L + 2z_b} \qquad (2)$$

This expression is verified in measurements of optical transmission vs. $L$, in which the incident angle is varied in samples which are and are not index matched to their surroundings (4).

We find an expression for the transmission averaged over random configurations and over all incident channels, $\langle T_a \rangle_a = \langle T \rangle / N = u(L)v_+ = W(L)$, by considering the energy density, $u(x)$, associate with the transmittance $\langle T \rangle$, as shown in Supplementary Figure 2. Here, $v_+$ is the average magnitude of the longitudinal speed of the wave in the medium. The simulations presented in Fig. 2c of the main text of the normalized energy density within a random medium shows that $\langle W(x) \rangle$ falls linearly within samples shorter than the localization length from $W(0)=2-\langle T \rangle/N$ at the incident surface to $W(L)=\langle T \rangle/N$ at the output of the sample.

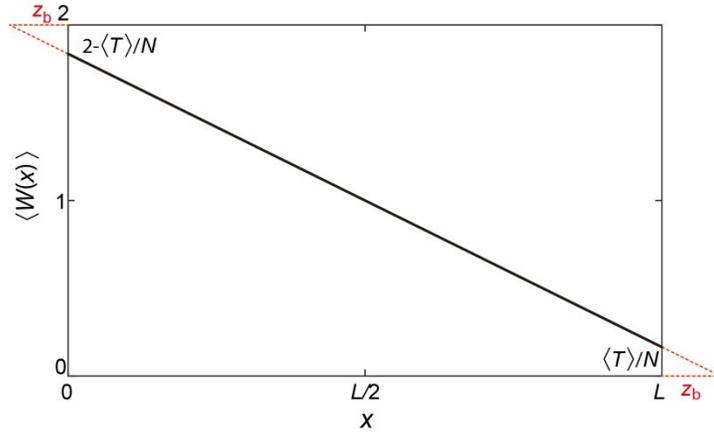

**Supplementary Figure 2| Linear falloff of $\langle W(x) \rangle$ within a scattering medium and its extrapolation beyond the sample.** $\langle W(x) \rangle$ extrapolates to 0 at $L+z_b$ and to 2 at $-z_b$.

Since the gradient of the average energy density is, $\dfrac{du(x)}{dx} = \dfrac{dW(x)}{v_+ dx} = -\dfrac{\langle T \rangle/N}{z_b v_+}$, the flux at the output $x = L$ is

$$\frac{\langle T \rangle}{N} = -\frac{v\ell}{2}\frac{du(x)}{dx} = \frac{v\ell \langle T \rangle/N}{2v_+ z_b}.$$

(3)

This gives

$$z_b = v\ell / 2v_+. \qquad (4)$$

Since $\langle W(x) \rangle$ extrapolates to zero at $x = L + z_b$ and to 2 at $x = -z_b$, $\dfrac{du(x)}{dx}$ can also be expressed as

$$\frac{du(x)}{dx} = -\frac{2}{(L+2z_b)v_+}.$$

(5)

The flux at the output is thus given by

$$\frac{\langle T \rangle}{N} = \frac{(v/v_+)\ell}{L+2z_b}.$$

(6)

This may be compared to the average of Supplementary Equation 2 over all incident channels,

$$\langle T_a \rangle_a = \frac{\langle T \rangle}{N} = \frac{\overline{z}_p + z_b}{L+2z_b}.$$

(7)

where $\overline{z}_p$ is the average of $z_p$ over all incident channels. This gives $\overline{z}_p + z_b = (v/v_+)\ell$ and

$$\overline{z}_p = z_b,$$ (8) in the case

that the sample is index matched and there is no reflection at the sample's longitudinal boundaries.

## Supplementary Note 2 - Universality of structure of $x_n$ for waves in random media

The $x_n$, which are related to the transmission eigenvalues via $\tau_n = 1/\cosh^2 x_n$ (5, 6), are seen in Fig. 3a of the main text to be equally spaced for $n < N/2$ in both translucent and diffusive samples. For opaque samples, $L \gg \ell$, the spacing between the $x_n$ for $n < N/2$ is predicted to be the inverse of the bare conductance, $\Delta x = 1/g_0 \sim L/N\ell$, in which the renormalization of the conductance by coherent backscattering and boundary effects are not included (5,6). Measurements in samples with $L$ not much larger than $\ell$, show that $\Delta x = (L+2z_b)/\eta N\ell$ with $\eta \sim 1$ (7). This suggests that when the effect of surface reflectivity is taken into account, the bare conductance is given by $g_0 = \eta N\ell/(L+2z_b)$. Since Supplementary Equation (4) and (6) yield $\langle T \rangle = Nz_b/(L+2z_b)$, this gives $\eta = z_b/\ell = v/2v_+$.

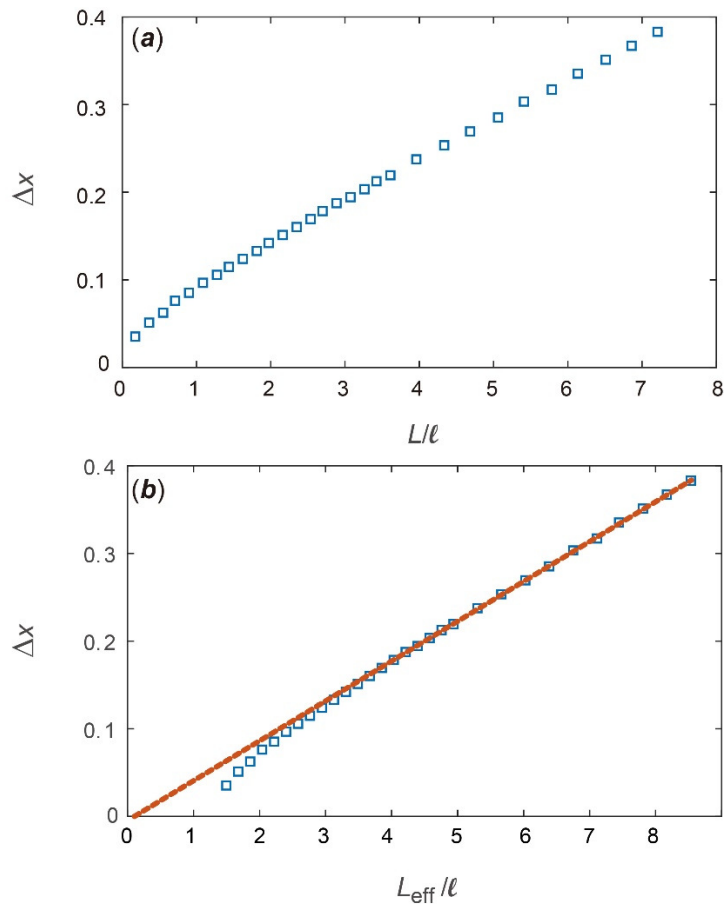

**Supplementary Figure 3| Scaling of $\Delta x$ vs $L/\ell$ and $L_{\text{eff}}/\ell$.** The red dashed line in Fig. 3b is a linear fit to the data for $L_{\text{eff}}/\ell > 3$ and is seen to intercept $\Delta x = 0$ approximately at $L_{\text{eff}} = 0$ or equivalently $L = -2z_b$.

For translucent samples, $\Delta x$ is not proportional to L, but varies linearly with L for $L > \ell$, as seen in Supplementary Fig. 3a. Supplementary Figure 3b shows that for $L > \ell$, $\Delta x$ is proportional to (L+2zb)/$\ell$ and extrapolates to zero at L = -2zb. Thus (L+2zb) is an effective length Leff, that is determined from the scaling of the xn.

The probability distribution of the $x$, $\rho(x,L)$, for various lengths normalized by $L/\ell$ or by $L_{\text{eff}}/\ell$ is shown in Supplementary Figs. 4a,b. For the two diffusive samples considered, $\rho(x,L)$ collapses to a single curve when normalized by $L_{\text{eff}}/\ell$, suggesting the universal distribution of $\rho(x,L)$ for opaque diffusive samples, $L \gg \ell$. The universal distribution breaks down, however, for translucent samples, $L \ll \ell$. The distribution $\rho(x,L)$ of $x$ normalized by $\Delta x$, is universal for both translucent and diffusive samples, as seen in Supplementary Fig. 4c.

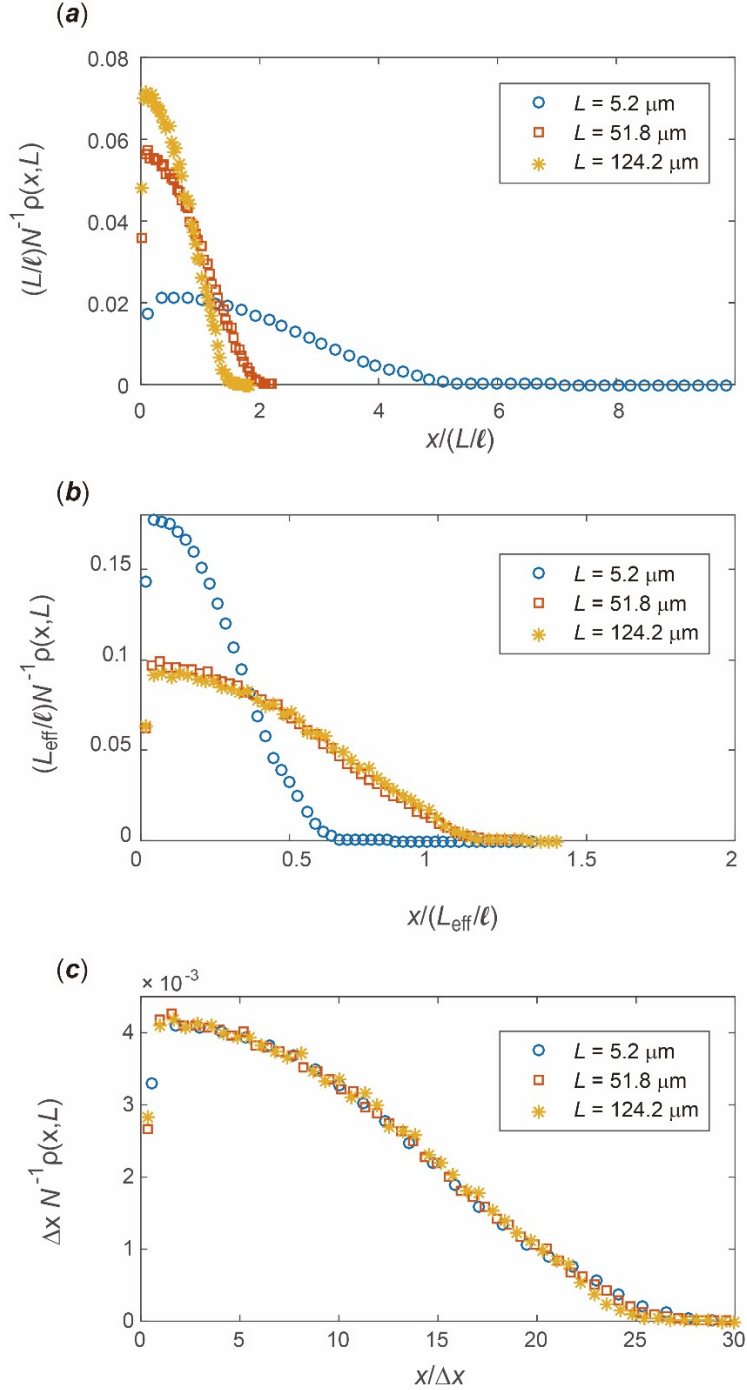

**Supplementary Figure 4| Universal structure of the density of $x_n$.** (a-c) Probability distribution of $\rho(x,L)$ plotted vs. $x/(L/\ell)$, $x/((L+2z_b)/\ell)$ and $x/\Delta x$, respectively.

**Supplementary Note 3 - Expression for $S_\tau(x/L)$ for diffusive and translucent media**

For an eigenchannel of a diffusing system with eigenvalue $\tau$, the expression for $S_\tau(x/L)$ is found in simulations to be

$$S_\tau(x/L) = 2\frac{\cosh^2(h(x/L)(1-x/L)(L/\xi'))}{\cosh^2(h(x/L)L/\xi')} - \tau, \qquad (9)$$

where $h(x/L)$ is an empirical function and $\tau = 1/\cosh^2(L/\xi')$ (8). $h(x/L)$ is found by comparing Supplementary Eq. (9) with the results of simulations for a specific value of $\tau$. With the function $h(x/L)$ obtained in this way, excellent agreement is found between simulation and Supplementary Eq. (9) in diffusive samples for all values of $\tau$. The need for the empirical function $h(x/L)$ for diffusive samples is seen in the comparisons in Supplementary Fig. 4 for a sample with $L = 124.2$ µm, of the simulation results, Supplementary Eq. (9), and the expression in Supplementary Eq. (9), but without the empirical function $h(x/L)$. This demonstrates the need for the empirical function $h(x/L)$ in diffusive samples. In contrast, as shown in Fig. 4c of the main text, good agreement is found between simulations and the expression in Supplementary Eq. (9) without the empirical function $h(x/L)$ in translucent samples. This is illustrated in Supplementary Fig. 5 for $\tau = 0.7$ in a sample with $L = 5.2$ µm.

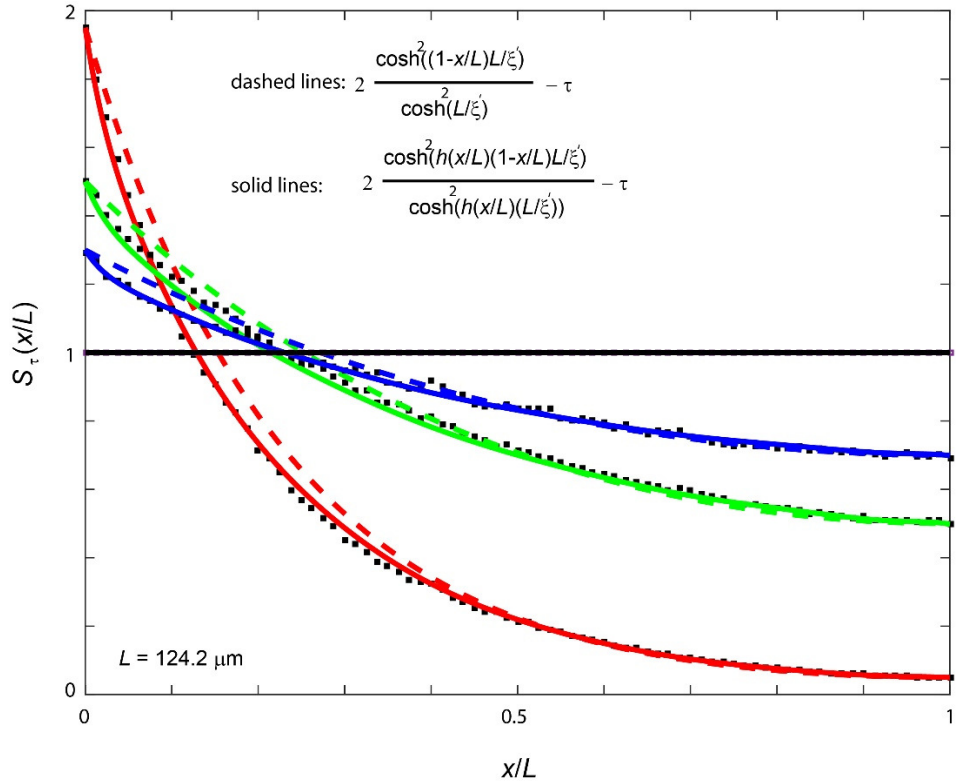

**Supplementary Figure 5|** Comparison of the simulations with the expression for $S_\tau(x/L)$ for various values of $\tau$ for a diffusing sample with $L = 124.2$ μm. The symbols are the points obtained in simulations for $\tau = 1, 0.7, 0.5$ and $0.01$, as indicted by the value of $S_\tau(1)=\tau$.

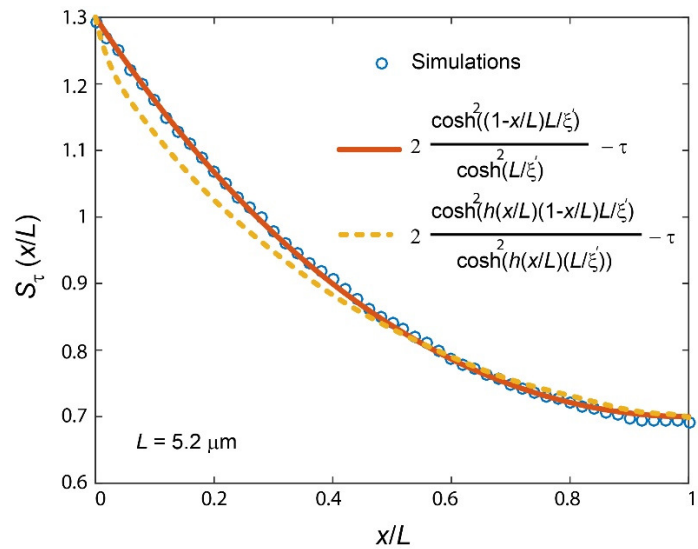

**Supplementary Figure 6|** $S_\tau(x/L)$ **in a translucent sample of $L = 5.2$ μm.** Comparison of the simulation results with expression for $S_\tau(x/L)$ for $\tau = 0.7$ in a sample with and without the empirical factor $h(x/L)$.

**Supplementary Note 4 - Scaling of the eigenchannel delay time of fully transmitting $t_1$ and the delay time $t_D$ and the scaling of $\tau_n$**

The transmission delay time $t_D$ is equal to the sum of the single channel delay time weighted by the intensity, $t_D = \sum_{a,b=1}^{N} |t_{ba}|^2 \frac{d\varphi_{ba}}{d\omega} / \sum_{a,b}^{N} |t_{ba}|^2$, where $\frac{d\varphi_{ba}}{d\omega}$ is the derivative of the phase of $t_{ba}$ with respect to the angular frequency $\omega$ (9-11). The delay time is also proportional to the density of states of the sample and to the sum of the energy stored within the medium for each of the eigenchannels (8,12-15). Alternatively, $t_D$ may be obtained from the sum of the differences in the derivative with angular frequency of the composite phase of the transmission eigenchannel on the outgoing and incoming surfaces of the sample summed over all transmission eigenchannels (15). The eigenchannel dwell time, $t_n$, is proportional to the contribution of the eigenchannel to the density of the states, and the energy stored within the sample in a transmission eigenchannel.

The delay time $t_D$ can also be expressed in terms of the $t_n$ and $\tau_n$, $t_D = \sum_1^N \tau_n t_n / \sum_1^N \tau_n$. The delay time of the eigenchannel is given by integrating the energy density distribution of the eigenchannels over the sample, $t_n \propto \int_0^L W_n(x) dx$. The same results are obtained when the eigenchannel delay time is obtained from the spectral derivative of the composite phase associated with the transmission eigenchannel, $t_n = \frac{d\theta_n}{d\omega}$, where $\frac{d\theta_n}{d\omega} = \frac{1}{i}(u_n^* \frac{du_n}{d\omega} - v_n^* \frac{dv_n}{d\omega})$ (8).

Since both $t_n$ and $\tau_n$ fall rapidly once $\tau_n <1/e$, $t_D$ is dominated by the $g$ open transmission eigenchannels. The scaling of various $\tau_n$ and $t_n/t_B$ are shown in Supplementary Figure 7.

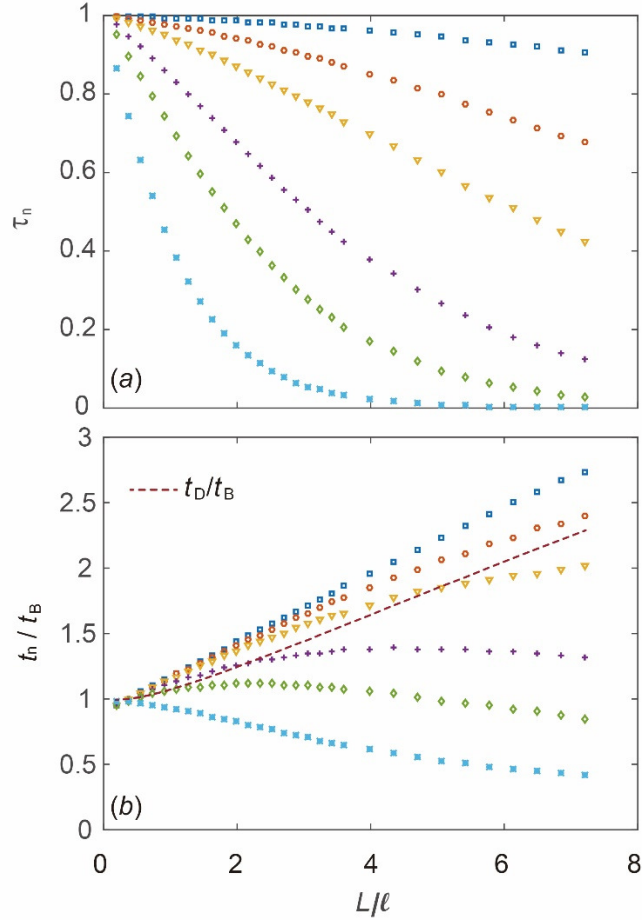

**Supplementary Figure 7| Scaling of transmission eigenchannels and dwell times.** Scaling of $\tau_n$ and $t_n/t_B$ for $n$ = 1, 2, 3, 5, 7 and 10 (from top to bottom) vs. $L/\ell$.

Computing the integral over the sample length $L$ of the energy density profile for the fully transmitting eigenchannel, $W_1(x)=1+F_1(x/L)=1+A[4(x/L)(1-x/L)]$, to give $t_1$ yields $t_1 \propto L+\frac{2}{3}AL$. For $L < \xi$, the value of $A$ can be found by considering the return probability at the center of the sample, $A = \frac{Lv_+}{4D}$ (8). For 2D samples, $D = v\ell/2$ and therefore $A = v_+L/2v\ell$. This gives

$$t_1 \propto L + \frac{v_+}{3v\ell}L^2$$

(10)

From Supplementary Eq. (10), it is clear that only when $L$ is substantially greater than the transport mean free path is the delay time of the highest and other high-transmission eigenchannels, $t_1$ and $t_\tau$, scale substantially faster than linearly.

**Supplementary Note 5 - Scaling of delay time $t_D$ with the effective length $L_{eff}$**

We have seen in measurements, simulations and calculations that transmission scales inversely with the effective length, $L_{eff} = L+2z_b$. We now consider the role of $L_{eff}$ in the scaling of dynamics. Supplementary Figure 8 shows that $t_D$ is proportional to $L_{eff}^2$ once $L_{eff} > 4\ell$. Thus the dynamics of diffusive transport depends on $z_b$ even though $z_b$ does not enter explicitly in the expressions for $\tau_n$ or $W_n(x)$.

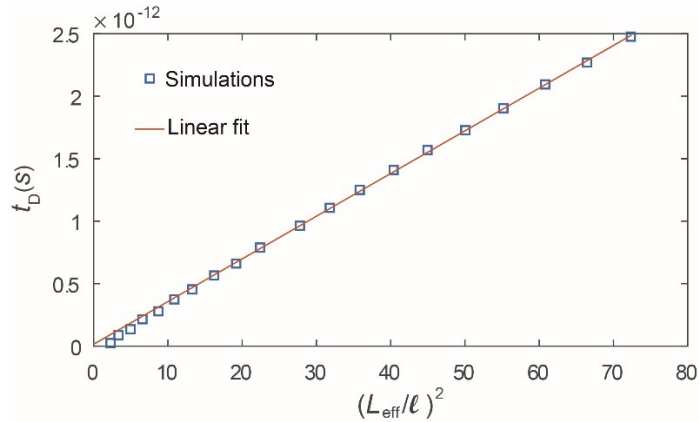

**Supplementary Figure 8| Scaling of $t_D$.** $t_D$ reflects the scaling of all the $t_n$ and $\tau_n$. It scales quadratically with $L_{eff}$ for $L_{eff}/\ell > 4$.

The quadratic scaling for larger lengths extrapolates to $t_D = 0$ at $L_{eff} = 0$. The linear fit is done for $(L_{eff}/\ell)^2 > 10$.

**Supplementary References**

1. Van Rossum, M. C. W. & Nieuwenhuizen, Th. M. Multiple scattering of classical waves: microscopy, mesoscopy, and diffusion, *Rev. Mod. Phys.* **71,** 313-371 (1999).

2. Akkermans, E. & Montambaux, G. *Mesoscopic physics of electrons and photons* (Cambridge University Press, 2007).

3. Zhu, J. Pine, D. J. & Weitz, D. A. Internal reflection of diffusive light in random media. *Phys. Rev. A* **44**, 3948-3959 (1991).

4. Li, J. H., Lisyansky, A. A., Cheung, T. D., Livdan, D. & Genack, A. Z. Transmission and surface intensity profiles in random media. *Europhys. Lett.* **22**, 675-680 (1993).

5. Dorokhov, O. N. On the coexistence of localized and extended electronic states in the metallic phase. *Solid State Commun.* **51**, 381–384 (1984).

6. Beenakker, C. W. J. Random-matrix theory of quantum transport, *Rev. Mod. Phys*. **69**, 731-808 (1997).

7. Shi, Z. & Genack, A. Z. Transmission eigenvalues and the bare conductance in the crossover to Anderson localization. *Phys. Rev. Lett.* **108**, 043901 (2012).

8. Davy, M., Shi, Z., Park, J., Tian, C. & Genack, A. Z. Universal structure of transmission eigenchannels inside opaque media. *Nature Commun.* **6**, 6893 (2015).

9. Genack, A. Z., Sebbah, P., Stoytchev, M., & van Tiggelen, B. A., Statistics of Wave Dynamics in Random Media, *Phys. Rev. Lett.* **82**, 715-718 (1999).

10. Wigner. E., Lower Limit for the Energy Derivative of the Scattering Phase Shift, *Phys. Rev.* **98**, 145-147 (1955).

11. Smith, F. T. Lifetime Matrix in Collision Theory, *Phys. Rev.* **118**, 349-356 (1960).

12. Avishai, Y. & Band, Y. One-dimensional density of states and the phase of the transmission amplitude. *Phys. Rev. B* **32**, 2674-2676 (1985).

13. Iannacone, G. General relation between density of states and dwell times in mesoscopic systems. *Phys. Rev. B* **51**, 4727–4729 (1995).

14. Brandbyge, M. & Tsukada, M. Local density of states from transmission amplitudes in multichannel systems, *Phys. Rev. B* **57**, R15088 (1998).

15. Davy, M, Shi, Z., Wang, J., Cheng, X. & Genack, A. Z. Transmission eigenchannels and the densities of states of random media. *Phys. Rev. Lett.* **114**, 033901 (2015).